\catcode`\@=11					



\font\fiverm=cmr5				
\font\fivemi=cmmi5				
\font\fivesy=cmsy5				
\font\fivebf=cmbx5				

\skewchar\fivemi='177
\skewchar\fivesy='60


\font\sixrm=cmr6				
\font\sixi=cmmi6				
\font\sixsy=cmsy6				
\font\sixbf=cmbx6				

\skewchar\sixi='177
\skewchar\sixsy='60


\font\sevenrm=cmr7				
\font\seveni=cmmi7				
\font\sevensy=cmsy7				
\font\sevenit=cmti7				
\font\sevenbf=cmbx7				

\skewchar\seveni='177
\skewchar\sevensy='60


\font\eightrm=cmr8				
\font\eighti=cmmi8				
\font\eightsy=cmsy8				
\font\eightit=cmti8				
\font\eightbf=cmbx8				

\skewchar\eighti='177
\skewchar\eightsy='60


\font\ninei=cmmi9
\font\ninesy=cmsy9

\skewchar\ninei='177
\skewchar\ninesy='60


\font\tenrm=cmr10				
\font\teni=cmmi10				
\font\tensy=cmsy10				
\font\tenex=cmex10				
\font\tenit=cmti10				
\font\tensl=cmsl10				
\font\tenbf=cmbx10				
\font\tentt=cmtt10				
\font\tenss=cmss10				
\font\tensc=cmcsc10				
\font\tenbi=cmmib10				

\skewchar\teni='177
\skewchar\tenbi='177
\skewchar\tensy='60

\def\tenpoint{\ifmmode\err@badsizechange\else
	\textfont0=\tenrm \scriptfont0=\sevenrm \scriptscriptfont0=\fiverm
	\textfont1=\teni  \scriptfont1=\seveni  \scriptscriptfont1=\fivemi
	\textfont2=\tensy \scriptfont2=\sevensy \scriptscriptfont2=\fivesy
	\textfont3=\tenex \scriptfont3=\tenex   \scriptscriptfont3=\tenex
	\textfont4=\tenit \scriptfont4=\sevenit \scriptscriptfont4=\sevenit
	\textfont5=\tensl
	\textfont6=\tenbf \scriptfont6=\sevenbf \scriptscriptfont6=\fivebf
	\textfont7=\tentt
	\textfont8=\tenbi \scriptfont8=\seveni  \scriptscriptfont8=\fivemi
	\def\rm{\tenrm\fam=0 }%
	\def\it{\tenit\fam=4 }%
	\def\sl{\tensl\fam=5 }%
	\def\bf{\tenbf\fam=6 }%
	\def\tt{\tentt\fam=7 }%
	\def\ss{\tenss}%
	\def\sc{\tensc}%
	\def\bmit{\fam=8 }%
	\rm\setparameters\setbaselines\fi}


\font\twelverm=cmr12				
\font\twelvei=cmmi12				
\font\twelvesy=cmsy10	scaled\magstep1		
\font\twelveex=cmex10	scaled\magstep1		
\font\twelveit=cmti12				
\font\twelvesl=cmsl12				
\font\twelvebf=cmbx12				
\font\twelvett=cmtt12				
\font\twelvess=cmss12				
\font\twelvesc=cmcsc10	scaled\magstep1		
\font\twelvebi=cmmib10	scaled\magstep1		

\skewchar\twelvei='177
\skewchar\twelvebi='177
\skewchar\twelvesy='60

\def\twelvepoint{\ifmmode\err@badsizechange\else
	\textfont0=\twelverm \scriptfont0=\eightrm \scriptscriptfont0=\sixrm
	\textfont1=\twelvei  \scriptfont1=\eighti  \scriptscriptfont1=\sixi
	\textfont2=\twelvesy \scriptfont2=\eightsy \scriptscriptfont2=\sixsy
	\textfont3=\twelveex \scriptfont3=\tenex   \scriptscriptfont3=\tenex
	\textfont4=\twelveit \scriptfont4=\eightit \scriptscriptfont4=\sevenit
	\textfont5=\twelvesl
	\textfont6=\twelvebf \scriptfont6=\eightbf \scriptscriptfont6=\sixbf
	\textfont7=\twelvett
	\textfont8=\twelvebi \scriptfont8=\eighti  \scriptscriptfont8=\sixi
	\def\rm{\twelverm\fam=0 }%
	\def\it{\twelveit\fam=4 }%
	\def\sl{\twelvesl\fam=5 }%
	\def\bf{\twelvebf\fam=6 }%
	\def\tt{\twelvett\fam=7 }%
	\def\ss{\twelvess}%
	\def\sc{\twelvesc}%
	\def\bmit{\fam=8 }%
	\rm\setparameters\setbaselines\fi}


\font\fourteenrm=cmr12	scaled\magstep1		
\font\fourteeni=cmmi12	scaled\magstep1		
\font\fourteensy=cmsy10	scaled\magstep2		
\font\fourteenex=cmex10	scaled\magstep2		
\font\fourteenit=cmti12	scaled\magstep1		
\font\fourteensl=cmsl12	scaled\magstep1		
\font\fourteenbf=cmbx12	scaled\magstep1		
\font\fourteentt=cmtt12	scaled\magstep1		
\font\fourteenss=cmss12	scaled\magstep1		
\font\fourteensc=cmcsc10 scaled\magstep2	
\font\fourteenbi=cmmib10 scaled\magstep2	

\skewchar\fourteeni='177
\skewchar\fourteenbi='177
\skewchar\fourteensy='60

\def\fourteenpoint{\ifmmode\err@badsizechange\else
	\textfont0=\fourteenrm \scriptfont0=\tenrm \scriptscriptfont0=\sevenrm
	\textfont1=\fourteeni  \scriptfont1=\teni  \scriptscriptfont1=\seveni
	\textfont2=\fourteensy \scriptfont2=\tensy \scriptscriptfont2=\sevensy
	\textfont3=\fourteenex \scriptfont3=\tenex \scriptscriptfont3=\tenex
	\textfont4=\fourteenit \scriptfont4=\tenit \scriptscriptfont4=\sevenit
	\textfont5=\fourteensl
	\textfont6=\fourteenbf \scriptfont6=\tenbf \scriptscriptfont6=\sevenbf
	\textfont7=\fourteentt
	\textfont8=\fourteenbi \scriptfont8=\tenbi \scriptscriptfont8=\seveni
	\def\rm{\fourteenrm\fam=0 }%
	\def\it{\fourteenit\fam=4 }%
	\def\sl{\fourteensl\fam=5 }%
	\def\bf{\fourteenbf\fam=6 }%
	\def\tt{\fourteentt\fam=7}%
	\def\ss{\fourteenss}%
	\def\sc{\fourteensc}%
	\def\bmit{\fam=8 }%
	\rm\setparameters\setbaselines\fi}


\font\seventeenrm=cmr10 scaled\magstep3		


\newdimen\rp@
\newcount\@basestretchnum
\newskip\@baseskip
\newskip\headskip
\newskip\footskip


\def\setparameters{\rp@=.1em
	\headskip=24\rp@
	\footskip=\headskip
	\delimitershortfall=5\rp@
	\nulldelimiterspace=1.2\rp@
	\scriptspace=0.5\rp@
	\abovedisplayskip=10\rp@ plus3\rp@ minus5\rp@
	\belowdisplayskip=10\rp@ plus3\rp@ minus5\rp@
	\abovedisplayshortskip=5\rp@ plus2\rp@ minus4\rp@
	\belowdisplayshortskip=10\rp@ plus3\rp@ minus5\rp@
	\normallineskip=\rp@
	\lineskip=\normallineskip
	\normallineskiplimit=0pt
	\lineskiplimit=\normallineskiplimit
	\jot=3\rp@
	\setbox0=\hbox{\the\textfont3 B}\p@renwd=\wd0
	\skip\footins=12\rp@ plus3\rp@ minus3\rp@
	\skip\topins=0pt plus0pt minus0pt}


\def\setbaselines{\maxdepth=4\rp@\baselinestretch=\@basestretchnum}


\def\baselinestretch{\afterassignment\@basestretch\@basestretchnum}
\def\@basestretch{%
	\@baseskip=12\rp@ \divide\@baseskip by1000
	\normalbaselineskip=\@basestretchnum\@baseskip
	\baselineskip=\normalbaselineskip
	\bigskipamount=\the\baselineskip
		plus.25\baselineskip minus.25\baselineskip
	\medskipamount=.5\baselineskip
		plus.125\baselineskip minus.125\baselineskip
	\smallskipamount=.25\baselineskip
		plus.0625\baselineskip minus.0625\baselineskip
	\setbox\strutbox=\hbox{\vrule height.708\baselineskip
		depth.292\baselineskip width0pt }}



\def\makeheadline{\vbox to0pt{\baselinestretch=1000
	\vskip-\headskip \vskip1.5pt
	\line{\vbox to\ht\strutbox{}\the\headline}\vss}\nointerlineskip}

\def\makefootline{\baselineskip=\footskip\line{\the\footline}}

\def\big#1{{\hbox{$\left#1\vbox to8.5\rp@ {}\right.\n@space$}}}
\def\Big#1{{\hbox{$\left#1\vbox to11.5\rp@ {}\right.\n@space$}}}
\def\bigg#1{{\hbox{$\left#1\vbox to14.5\rp@ {}\right.\n@space$}}}
\def\Bigg#1{{\hbox{$\left#1\vbox to17.5\rp@ {}\right.\n@space$}}}


\mathchardef\alpha="710B
\mathchardef\beta="710C
\mathchardef\gamma="710D
\mathchardef\delta="710E
\mathchardef\epsilon="710F
\mathchardef\zeta="7110
\mathchardef\eta="7111
\mathchardef\theta="7112
\mathchardef\iota="7113
\mathchardef\kappa="7114
\mathchardef\lambda="7115
\mathchardef\mu="7116
\mathchardef\nu="7117
\mathchardef\xi="7118
\mathchardef\pi="7119
\mathchardef\rho="711A
\mathchardef\sigma="711B
\mathchardef\tau="711C
\mathchardef\upsilon="711D
\mathchardef\phi="711E
\mathchardef\chi="711F
\mathchardef\psi="7120
\mathchardef\omega="7121
\mathchardef\varepsilon="7122
\mathchardef\vartheta="7123
\mathchardef\varpi="7124
\mathchardef\varrho="7125
\mathchardef\varsigma="7126
\mathchardef\varphi="7127
\mathchardef\imath="717B
\mathchardef\jmath="717C
\mathchardef\ell="7160
\mathchardef\wp="717D
\mathchardef\partial="7140
\mathchardef\flat="715B
\mathchardef\natural="715C
\mathchardef\sharp="715D


\def\err@badsizechange{%
	\immediate\write16{--> Size change not allowed in math mode, ignored}}

\baselinestretch=1000
\tenpoint

\catcode`\@=12					
\catcode`\@=11
\expandafter\ifx\csname @iasmacros\endcsname\relax
	\global\let\@iasmacros=\par
\else	\immediate\write16{}
	\immediate\write16{Warning:}
	\immediate\write16{You have tried to input iasmacros more than once.}
	\immediate\write16{}
	\endinput
\fi
\catcode`\@=12


\def\rmb{\seventeenrm}

\def\singlespace{\baselineskip=\normalbaselineskip}
\def\halfspace{\baselineskip=1.5\normalbaselineskip}
\def\doublespace{\baselineskip=2\normalbaselineskip}


\def\AB{\bigskip\parindent=40pt
        \centerline{\bf ABSTRACT}\medskip\halfspace\narrower}
\def\AE{\bigskip\nonarrower\doublespace}
\def\nonarrower{\advance\leftskip by-\parindent
	\advance\rightskip by-\parindent}


\def\boxit#1{\vbox{\hrule\hbox{\vrule\kern3pt
	\vbox{\kern3pt#1\kern3pt}\kern3pt\vrule}\hrule}}

\def\hence{\leavevmode\hbox{\bf .\raise5.5pt\hbox{.}.} }

\def\dalemb#1#2{{\vbox{\hrule height.#2pt
	\hbox{\vrule width.#2pt height#1pt \kern#1pt \vrule width.#2pt}
	\hrule height.#2pt}}}
\def\gtorder{\mathrel{\raise.3ex\hbox{$>$}\mkern-14mu
             \lower0.6ex\hbox{$\sim$}}}
\def\ltorder{\mathrel{\raise.3ex\hbox{$<$}\mkern-14mu
             \lower0.6ex\hbox{$\sim$}}}

\newdimen\fullhsize
\newbox\leftcolumn
\def\twoup{\hoffset=-.5in \voffset=-.25in
  \hsize=4.75in \fullhsize=10in \vsize=6.9in
  \def\fullline{\hbox to\fullhsize}
  \let\lr=L
  \output={\if L\lr
        \global\setbox\leftcolumn=\columnbox\global\let\lr=R \advancepageno
      \else \doubleformat \global\let\lr=L\fi
    \ifnum\outputpenalty>-20000 \else\dosupereject\fi}
  \def\doubleformat{\shipout\vbox{
    \fullline{\box\leftcolumn\hfil\columnbox}\advancepageno}}
  \def\columnbox{\leftline{\vbox{\makeheadline\pagebody\makefootline}}}
  \tolerance=1000 }

\twelvepoint
\overfullrule=0pt
{\nopagenumbers{
\rightline{~~~June, 2003}
\bigskip\bigskip
\centerline{\rmb Global unitary fixing and matrix-valued correlations in
matrix models}
\bigskip\bigskip
\bigskip
\medskip                        
\centerline{\rm Stephen L. Adler \footnote{$^a$}{\twelvepoint 
Electronic mail: 
adler@ias.edu} 
and Lawrence P. Horwitz \footnote{$^b$}{\singlespace  On 
leave from the School of Physics  
and Astronomy, Raymond and Beverly Sackler Faculty of Exact Sciences, 
Tel Aviv University, Israel, and Department of Physics, Bar Ilan University, 
Ramat Gan, Israel} } 
\centerline{\it Institute for Advanced Study}
\centerline{\it Princeton, NJ 08540, USA}
\medskip

\bigskip\bigskip
\bigskip\bigskip
\leftline{\it Send correspondence to:}
\medskip
{\singlespace\leftline{Stephen L. Adler}
\leftline{Institute for Advanced Study}
\leftline{Einstein Drive, Princeton, NJ 08540}
\leftline{Phone 609-734-8051; FAX 609-924-8399; email adler@ias.
edu}}
\bigskip\bigskip
}}
\vfill\eject
\pageno=2
\doublespace     
\AB
We consider the partition function for a matrix model 
with a global unitary invariant energy function.  We show that the 
averages over the partition function of global unitary invariant trace 
polynomials of the matrix variables are the same when calculated with any 
choice of a global unitary fixing, while averages of 
such polynomials without a  
trace define matrix-valued correlation functions, that depend on the 
choice of unitary fixing.  The unitary fixing is  
formulated within the standard Faddeev-Popov framework, in which the squared 
Vandermonde determinant emerges as a factor of the complete Faddeev-Popov 
determinant.  
We give the ghost representation for the FP determinant, and 
the corresponding 
BRST invariance of the unitary-fixed partition function.  The formalism is 
relevant for deriving Ward identities obeyed by matrix-valued correlation 
functions.  
\AE
\bigskip\bigskip
\vfill\eject
\pageno=3
Over the years there has been considerable interest in matrix models from 
various points of view.  Matrix models are used to approximate quantum many 
body systems and quantum field 
theories [1], and have deep connections with string theories [2].  
They also have 
been studied as classical statistical mechanical systems, from which quantum 
behavior emerges under certain conditions [3].  A common issue 
that arises in 
all of these applications is dealing with an overall global unitary 
invariance transformation of the partition function.  Typically, in matrix  
model calculations this overall invariance is partially 
integrated out as a first step, thus  
eliminating a $U(N)/U(1)^N$ subgroup of the global unitary group.  
Our aim in this paper is to proceed in an alternative fashion, by using 
the Faddeev-Popov framework to impose a set of unitary invariance 
fixing conditions, that completely break the $SU(N)$ subgroup of the 
global unitary invariance group $U(N)$.  One can think of our construction     
as a type of polar decomposition, based on modding out the action of the 
$SU(N)$ subgroup. 
This allows one to define matrix-valued correlation functions, which give 
additional structural information about the system, but which (like gauge 
potentials in gauge field theory) depend on the choice of unitary fixing.  
A complete global unitary fixing is needed for the application of matrix 
models to emergent quantum theory developed in Ref. [3], so as to be able  
to construct matrix ensembles that do not integrate over the spacetime translation group of the emergent  
theory.  The formalism that we develop here 
may well find other matrix model applications as well. 

Let $M_1,..., M_D$ be a set of $N\times N$ complex self-adjoint   
matrices, and let 
us take as the energy function 
$${\bf H}[\{M\}]={\rm Tr}H(M_1,..., M_D)~~~,\eqno(1)$$ 
with $H$ a self-adjoint polynomial in its arguments constructed using only 
$c$-number coefficients (i.e., no fixed, non-dynamical matrices appear as 
coefficients in constructing $H$).  Then the corresponding partition 
function $Z$ is defined by 
$$Z=\int dM \exp(-{\bf H})~~~,\eqno(2)$$
with 
$$dM = \prod_{d=1}^D d[M_d]~~~,\eqno(3a)$$ 
and with the integration measure $d[M]$ for the self-adjoint matrix $M$ 
defined in terms of the real and imaginary parts of the 
matrix elements $M_{ij}$ of $M$ by 
$$d[M]= \prod_i dM_{ii} \prod_{i<j} d{\rm Re} M_{ij} d{\rm Im} M_{ij}~~~.
\eqno(3b)$$
As is well known, the measure $d[M]$ is unitary invariant, in 
other words, if $U$ is a fixed $N \times N$ unitary matrix, then  
$$d[U^{\dagger}MU]=d[M]~~~.\eqno(4a)$$  
If we make the same unitary transformation $U$ 
on all of the matrices $M_d~,~~d=1,...,D$, then by our assumption that $H$  
involves no fixed matrix coefficients, ${\bf H}$ is invariant by 
virtue of the cyclic property of the trace,  
$$ {\bf H}[\{U^{\dagger}M U\}]= {\bf H}[\{M\}]~~~.\eqno(4b)$$ 
Thus, Eqs.~(4a) and (4b) together imply that the partition function $Z$ has 
a global unitary invariance.  

The global unitary invariance of $Z$ must be taken into account in 
calculating  correlations of the various matrices $M_d$ averaged over the 
partition function.  Let $Q[\{M\}]$ be an arbitrary polynomial in the  
matrices $M_1,..., M_D$ constructed using only $c$-number coefficients, 
so that under global unitary transformations, $Q$ transforms as 
$$Q[\{U^{\dagger}MU\}]= U^{\dagger} Q[\{M\}]  U ~~~.\eqno(5a)$$
Correspondingly, let 
$${\bf Q}={\rm Tr}Q~~~,\eqno(5b)$$ 
so that ${\bf Q}$ is a global 
unitary invariant.  One can now consider the calculation of averages of 
${\bf Q}$ and of $Q$ respectively over the ensemble defined by Eq.~(2).  
In the case of the trace polynomial ${\bf Q}$ one has 
$$ \langle {\bf Q} \rangle_{\rm AV}=
  Z^{-1}\int dM \exp(-{\bf H}) {\bf Q}~~~, \eqno(6a)$$   
which because of the global unitary invariance involves an overall 
structure-independent unitary 
integration that is typically done as the first step, by using Mehta's 
change of variables [4] for one of the matrix arguments on which $Q$ depends. 
Let us now consider the corresponding average of the polynomial $Q$ 
over the ensemble, 
 $$ \langle  Q \rangle_{\rm AV}=
  Z^{-1}\int dM \exp(-{\bf H})  Q~~~. \eqno(6b)$$   
Making a global unitary transformation on all of the matrix integration 
variables, and using the invariance of $dM$  and of ${\bf H}$ given 
in Eqs.~(4a,b), 
and the covariance of $Q$ given in Eq.~(5a), we then find that 
$$\langle Q \rangle_{\rm AV} = U^{\dagger} \langle Q \rangle_{\rm AV} U~~~,
\eqno(7a)$$
for all unitary matrices $U$.  Thus by Schur's lemma (which applies since 
$U(N)$ acts irreducibly on the complex $N$ dimensional vector space)  
$ \langle Q \rangle_{\rm AV} $ must be a $c$-number multiple of the 
unit matrix, 
so that by taking the trace, we learn that 
$$\langle Q \rangle_{\rm AV} =N^{-1} \langle {\bf Q} \rangle_{\rm AV}~~~,
\eqno(7b)$$
and all nontrivial matrix information (e.g., the unitary 
orientation and nontrivial operator properties) contained in  $Q$ 
has been lost.  

In order to retain access to the matrix information contained in  
$Q$, let us then proceed in an alternative fashion.  
Let us define a measure $\hat dM$ 
in which the $SU(N)$ subgroup of the 
global unitary invariance group has been fixed. 
(The full global unitary 
invariance group is the product of this $SU(N)$ with a global $U(1)$ that 
is an overall phase times the unit matrix;  since this $U(1)$ commutes 
with $Q$, averaging over it causes no loss of the matrix information 
contained in $Q$, and so fixing the overall $U(1)$ is not necessary.) 
We then define the  
average of $Q$ over the unitary-fixed ensemble as 
$$\langle Q \rangle_{\hat{\rm AV}} =
\hat Z^{-1} \int \hat dM \exp(-{\bf H}) Q~~~, \eqno(8a)$$   
with 
$$ \hat Z = \int \hat dM \exp(-{\bf H})~~~\eqno(8b)$$ 
the partition function in which the global unitary invariance 
has been broken, and an orientation on the 
$N$-dimensional vector space has been fixed.  
Clearly, the procedure just described is a global unitary 
analog of the gauge fixing customarily employed in the case of local gauge 
invariances.  If we change the recipe for fixing the global unitary 
invariance, then the average defined by Eq.~(8a) will change in a 
manner that is in general complicated.  However, we will show that the 
average of ${\bf Q}$ in the unitary-fixed ensemble  
is independent of the fixing and  
is equal to that defined in Eq.~(6a) by averaging over the original ensemble, 
so that 
$$\langle {\bf Q} \rangle_{\hat{\rm AV}}=\langle {\bf Q} \rangle_{\rm AV}~~~.
\eqno(9)$$
In other words, the average of the trace of $Q$ takes the same value for  
any choice of unitary fixing.  To make an analogy with local gauge fixing 
in gauge theories, the trace polynomials ${\bf Q}$ are analogs of 
gauge invariant functions, while polynomials $Q$ without a trace are analogs 
of gauge-variant quantities.  Just as the gauge-variant potentials contain 
useful information in gauge theories, the unitary fixing-variant averages of 
polynomials $Q$ contain useful structural information about matrix models.  

To prove Eq.~(9), we proceed by analogy with the standard Faddeev-Popov 
procedure used for local gauge fixing.  Let us write an infinitesimal    
$SU(N)$ transformation in generator form as $U=\exp(G)$, with $G$ 
anti-self-adjoint and traceless.  We take as the $N^2-1$ infinitesimal 
parameters 
of the $SU(N)$ transformation the real numbers $g_j~,~~j=1,..., N^2-1$, with 
those for $j=1,..., N(N-1)$ given by the real and imaginary parts of  
the off-diagonal matrix elements of $G$, that is, by ${\rm Re}\,G_{ij}$ 
and ${\rm Im}\,G_{ij}$ for $i<j$. The remaining ones for $j=N(N-1)+1,...,
N^2-1$ are given by the differences of 
the imaginary parts of the 
diagonal matrix elements of $G$, that is, by 
${\rm Im} (G_{11}-G_{22}),...,{\rm Im}(G_{11}-G_{NN})$.  
Let $f_j(\{M\})~,~j=1,...,N^2-1$ be a set of functions of the matrices 
$M_1,...,M_D$ with the property that the equations 
$f_j(\{M\})=0~,~j=1,...,N^2-1$ completely break the $SU(N)$ invariance group, 
so that the only 
solution of $f_j(\{M+[G,M]\})=0~,~j=1,...,N^2-1$ is $g_j=0~,~j=1,...,N^2-1$.
We consider now the integral  
$${\cal J}=\int dM {\cal G}[\{M\}] K[\{f_j\}]  
{\rm det}\left( {\partial f_i(\{M+[G,M]\}) \over \partial g_j }\Big|_{G=0} 
\right)~~~,\eqno(10a)$$ 
with the function $ K[\{f_j\}]$ taken as   
$$K[\{f_j\}]=\prod_{j=1}^{N^2-1}\delta(f_j)~~~.\eqno(10b)$$    
Here ${\cal G}$ is a global unitary invariant function of the matrices 
$M_1,..., M_D$, such as a trace polynomial ${\bf Q}$ or any function 
of trace polynomials (for example the 
partition function weight $\exp(-{\bf H})$).  
Equation (10a) has the standard form of the Faddeev-Popov 
analysis, as formulated for example 
in the text of Weinberg [5] (except that when one   
is dealing with a non-compact local gauge invariance, where the limits of 
integration lie at infinity, one can take the 
function $K$ to be a general function of gauge variant functions $f_j$; in 
the compact case considered here, the delta functions of Eq.~(10b) must be 
used in order to make the integration limits irrelevant.)  
The standard FP argument then shows that the integral 
in Eq.~(10a) is independent of the constraints $f_j$. Briefly, the argument 
proceeds by replacing the dummy variable of integration $dM$ by $dM^V$, where 
$M^V=V^{\dagger} M V$, and integrating over the $SU(N)$ matrix $V$.  The 
group property of unitary transformations together with the chain rule then  
converts the determinant in Eq.~(10a) into a Jacobian transforming the 
$V$ integration into an integration over the constraints $f_j$, permitting 
the delta functions in Eq.~(10b) to be integrated to give unity. This shows   
that the result is independent of the constraints, 
and that it is the same as the    
result obtained by integrating over the original unfixed ensemble, 
thus establishing Eq.~(9).  Clearly, this argument works 
only when the function ${\cal G}$ is a unitary invariant, so that it has 
no dependence on $V$.  For example, if ${\cal G}$ is replaced by a polynomial 
in the matrices  {\it without} an overall trace, then the unitary fixing  
constraints cannot be 
eliminated by integrating over $V$, and the result depends on the unitary 
fixing in a complicated way.  

A specific realization of the general unitary fixing can be given when 
$D \geq 2$, so that the set of matrices $M_1,...,M_D$ contains at least 
two independent self-adjoint matrices $A=M_1$ and $B=M_2$.  We take 
the functions 
$f_j~,~j=1,...,N^2-1$ to be linear functions of $A$ and $B$, constructed 
as follows. As the $f_j$ for  
$j=1,..., N(N-1)$ we take the real and imaginary parts of  
the off-diagonal matrix elements of $A$, that is, the functions 
${\rm Re}\,A_{ij}$ 
and ${\rm Im}\,A_{ij}$ for $i<j$. Equating these functions to zero forces the 
matrix $A$ to be diagonal. The $N$ remaining diagonal unitary transformations 
then commute with $A$, so that no further conditions can be 
furnished by use of $A$ alone.  However, the diagonal $SU(N)$ transformations 
can always be used 
to make the off-diagonal matrix elements in the first row of the second  
matrix $B$ have vanishing imaginary parts, leaving a 
residual $Z_2^{N-1}$   
symmetry that is broken by requiring these matrix elements
 to have positive semidefinite real parts.
 So for the 
remaining conditions 
$f_j$ for $j=N(N-1)+1,...,N^2-1$, 
we take the $N-1$ functions ${\rm Im}B_{1j}~,~j>1$, and we restrict the 
integrations over ${\rm Re}B_{1j}~,~j>1$ to run from 0 to $\infty$.
  Since the   
function $K$ chosen in Eq.~(10b) enforces the conditions $f_j=0$ 
in a sharp manner, they   
can be used to simplify the expression for the Faddeev-Popov determinant. 
A simple calculation now shows that when the $f_j$  all vanish, the matrix 
elements of the commutator $[G,M]$ needed in Eq.~(10a) are given by 
$$\eqalign{
{\rm Re}[G,A]_{ij}=&{\rm Re}G_{ij}(A_{jj}-A_{ii})~~~,\cr 
{\rm Im}[G,A]_{ij}=&{\rm Im}G_{ij}(A_{jj}-A_{ii})~~~,\cr 
{\rm Im}[G,B]_{1j}=&{\rm Re}B_{1j} {\rm Im}(G_{11}-G_{jj}) +R~~~,\cr
}\eqno(11a)$$
with $R$ a remainder containing only off-diagonal elements $G_{i\not= j}$ 
of the matrix $G$.  
Since Eq.~(11a) shows that the matrix  
$$\left( {\partial f_i(\{M+[G,M]\}) \over \partial g_j }\Big|_{G=0} 
\right)~~~\eqno(11b)$$ 
is triangular (its upper off-diagonal matrix elements are all zero because 
$R$ has no dependence on the diagonal matrix elements of $G$), its 
determinant is given by the product of its diagonal matrix elements.  
Thus we have  
$$\Delta \equiv {\rm det}\left( {\partial f_i(\{M+[G,M]\}) 
\over \partial g_j }\Big|_{G=0} 
\right)~~~=\prod_{i<j} (A_{ii}-A_{jj})^2 \prod_{j=2}^N {\rm Re}B_{1j}~~~,
\eqno(12a)$$
the first factor of which is the familiar squared Vandermonde determinant.  
Substituting Eqs.~(10b) and (12a) into Eq.~(10a), we thus arrive 
at the formula for the unitary-fixed integral 
$${\cal J}=\int \prod_{d=3}^D  d[M_d]  (\prod_{i=1}^N dA_{ii} dB_{ii})
(\prod_{j=2}^N d{\rm Re}B_{1j}) 
(\prod_{2 \leq i<j} d{\rm Re}B_{ij} d{\rm Im}B_{ij}) \Delta  
{\cal G}[\{M\}]~~~,\eqno(12b)$$ 
with the integrals over ${\rm Re}B_{1j}~,~j=2,...,N$
 in Eq.~(12b) running over positive values only.    
The part of this analysis involving only a single matrix $A$ is well-known 
in the literature [6]; what has been added here is the complete   
$SU(N)$ fixing obtained by imposing a condition on a 
second matrix $B$ as well.  The part of Eqs.~(12a,b) involving each $B_{1j}$ 
is just a planar radial integral $\int_0^{\infty} \rho d\rho$, with 
$\rho=|B_{1j}|={\rm Re} B_{1j}$, where the associated angular integral  
$\int_0^{2 \pi}d\phi$ has been omitted because it corresponds to a $U(1)$ 
factor that has been fixed by the condition $\phi=0$.  

With this choice of unitary fixing, the unitary fixed average 
$\hat {\bar Q}\equiv \langle Q \rangle_{\hat{\rm AV}}$ defined in Eq.~(8a)  
has a characteristic form that is dictated by the symmetries 
of the unitary-fixed ensemble.  
Since the unitary fixing conditions are symmetric 
under permutation of the basis states with labels $2,3,...,N$, and since  
this permutation is also a symmetry of the unfixed measure $dM$, the 
matrix $\hat{\bar Q}$ 
must be symmetric under this permutation of basis states.  Thus, there  
are only five independent matrix elements, 
$$\eqalign{
\hat{\bar Q}_{11}=&\alpha~~~,\cr
\hat{\bar Q}_{jj}=&\beta~,~j=2,...,N~~~,\cr
\hat{\bar Q}_{1j}=&\gamma~,~j=2,...,N~~~,\cr
\hat{\bar Q}_{i1}=&\delta~,~i=2,...,N~~~,\cr
\hat{\bar Q}_{ij}=&\epsilon~,~2\leq i\not =j \leq N~~~.\cr
}\eqno(12c)$$
In this notation, the original unfixed average 
$\bar Q\equiv \langle Q \rangle_{\rm AV}$
defined by Eq.~(7b) is  
given by 
$$\bar Q=N^{-1} {\rm Tr} \hat {\bar Q}
=N^{-1} [\alpha + (N-1) \beta]~~~,\eqno(12d)$$ 
showing explicitly that there is a loss of structural 
information in using the unfixed average.  But even the unitary-fixed 
average has a structure that is greatly restricted as compared with a 
general $N \times N$ matrix.  (Similar reasoning applies to the partial  
unitary fixing in which one only imposes the condition that $A$ should 
be diagonal.  Since this condition is symmetric under permutation of the 
basis states with labels $1,...,N$, the partially unitary fixed average of 
a polynomial $Q$ defined by integrating with the measure 
$(\prod_{i=1}^N dA_{ii}) \big(\prod_{i<j}(A_{ii}-A_{jj})^2\big) 
\prod_{d=2}^D d[M_d]$ 
must also have this permutation symmetry, and thus must be a $c$-number 
times the unit matrix.)  

We now introduce ghost integrals to represent the determinant $\Delta$. 
Let $\omega_{ij}$ and $\tilde \omega_{ij}$ be the matrix elements of 
independent $N \times N$ complex 
anti-self-adjoint Grassmann matrices $\omega$ and $\tilde \omega$. 
We take  $\omega$ to be traceless, ${\rm Tr}\omega=0$, while we take 
$\tilde \omega$ to have a vanishing 11 matrix element, 
$\tilde \omega_{11}=0$.  
The integration measure for $\omega$ is defined by    
$$d\omega = \prod_{i<j} d{\rm Re}\omega_{ij} d{\rm Im}\omega_{ij} 
\prod_{j=2}^N d{\rm Im}(\omega_{jj}-\omega_{11})~~~,\eqno(13a)$$ 
while the integration measure for $\tilde \omega$ is taken as 
$$d\tilde \omega = \prod_{i<j} d{\rm Re}\tilde \omega_{ij} 
d{\rm Im}\tilde \omega_{ij} 
\prod_{j=2}^N d{\rm Im}\tilde \omega_{jj}~~~.\eqno(13b)$$ 
We can now use these Grassmann matrices to give a ghost representation of the
factors in Eq.~(12a) involving the matrices $A$ and $B$.  
Since the matrix $A$ is diagonal, we have 
$${\rm Tr} \tilde \omega[\omega,A]=\sum_{i\not=j} 
\tilde\omega_{ji}(A_{ii}-A_{jj})
\omega_{ij}~~~.\eqno(14a)$$ 
Hence up to an overall sign, the square of the Vandermonde determinant  
$\prod_{i<j} (A_{ii}-A_{jj})^2$ is given by the ghost integral 
$$\int d^{\prime}\omega \, d^{\prime}\tilde \omega 
\exp({\rm Tr} \tilde \omega[\omega,A] )
~~~,\eqno(14b)$$
with the diagonal factors $d{\rm Im}(\omega_{jj}-\omega_{11}),\,
d{\rm Im}\tilde \omega_{jj}~,~~j=2,...,N$ 
omitted from the 
primed integration measures $d^{\prime}\omega$ and $d^{\prime}\tilde \omega$.  
To represent the second factor in Eq.~(12a) as a ghost integral, we 
use the diagonal matrix elements of $\omega$ and $\tilde \omega$ 
in an analogous fashion.  Thus, up to a 
phase, the factor $\prod_{j=2}^{N} {\rm Re}B_{1j}$ is given by 
the ghost integral 
$$\int \prod_{j=2}^N d{\rm Im}(\omega_{jj}-\omega_{11}) 
d{\rm Im}\tilde \omega_{jj}
\exp\big(\sum_{j=2}^N \tilde \omega_{jj} 
({\rm Re}B_{1j})i (\omega_{jj}-\omega_{11})\big)~~~.
\eqno(14c)$$
By defining a matrix 
$X$ by $X_{11}=0$; $X_{ij}=0~,~2\leq i,j \leq N$; 
$X_{1j}=X_{j1}={i\over 2} 
\tilde \omega_{jj}(\omega_{jj}-\omega_{11})~,~j=2,...,N$, 
the exponent in Eq.~(14c) 
can be written  
as ${\rm Tr}XB$, so that Eq.~(14c) becomes 
$$\int \prod_{j=2}^N d{\rm Im}(\omega_{jj}-\omega_{11}) 
d{\rm Im}\tilde \omega_{jj} 
\exp({\rm Tr}XB)~~~.\eqno(14d)$$
Combining Eqs.~(13a,b) and (14b,c,d), we see that up to an overall 
phase the determinant $\Delta$ introduced in Eq.~(12a) has the equivalent  
ghost representations 
$$\eqalign{
\Delta 
\propto&\int d\omega d\tilde \omega  
\exp\big({\rm Tr} \tilde \omega[\omega,A] + \sum_{j=2}^N 
\tilde \omega_{jj}({\rm Re} B_{1j})
i(\omega_{jj}-\omega_{11})\big)\cr
\propto& \int d\omega d\tilde \omega  
\exp({\rm Tr} \tilde \omega[\omega,A] + {\rm Tr} XB)~~~.\cr
}\eqno(15)$$

Yet another equivalent form is obtained by noting that 
$$[B,\omega]_{1j}=B_{1j}(\omega_{jj}-\omega_{11}) +S~~~,\eqno(16a)$$
with the remainder $S$ denoting terms that only involve matrix elements 
$\omega_{ij}$ with $i \not=j$.  The remainder $S$ makes a vanishing 
contribution to the Grassmann integrals when Eq.~(16a) is substituted for 
$B_{1j}i(\omega_{jj}-\omega_{11})$ in Eq.~(15), since one factor 
of $(\omega_{jj}-\omega_{11})$ for each $j=2,...,N$ is needed to give 
a nonvanishing integral, and each such term in the exponent is already 
accompanied by a factor $\tilde \omega_{jj}$, so that  
terms with additional such factors vanish inside the Grassmann integrals.  
(We are just using here the fact that with $\zeta,~\tilde \zeta$ Grassmann 
variables, $\int d\zeta d\tilde \zeta 
\exp(\tilde \zeta W \zeta + U\zeta) = W$, with no dependence on $U$.)
Since the diagonal matrix elements 
of $\omega$ are pure imaginary, Eq.~(16a) implies that 
$$({\rm Re} B_{1j})i(\omega_{jj}-\omega_{11})
=-{\rm Im} [B,\omega]_{1j}+{\rm Im} S~~~,\eqno(16b)$$
which when substituted into Eq.~(15) gives the alternative formula 
$$\Delta 
\propto \int d\omega d\tilde \omega  
\exp\big({\rm Tr} \tilde \omega[\omega,A] - \sum_{j=2}^N 
\tilde \omega_{jj} {\rm Im} [B,\omega]_{1j} \big)~~~.\eqno(17)$$
This formula will be used to establish a BRST [7] symmetry, 
the topic to which we now turn. 
 
To formulate a BRST invariance transformation corresponding to    
Eq.~(17), we rewrite the product of $\delta$ functions in Eq.~(10b) and 
the half-line restriction on the integrals over ${\rm Re}B_{1j}$ 
in terms of their Fourier representations, by introducing three sets of 
Nakanishi-Lautrup [8] variables.  One set are the elements $h_{ij}$ 
of a self-adjoint $N \times N$ matrix $h$ with vanishing diagonal 
matrix elements, so that $h_{ii}=0~,~i=1,...,N$.  The integration measure 
for this set is defined as 
$$dh=\prod_{i<j} d{\rm Re}h_{ij}d{\rm Im}h_{ij}~~~.\eqno(18a)$$  
The second set are $N-1$ real numbers $H_j~,~j=2,...,N$, with integration 
measure 
$$dH=\prod_{j=2}^N dH_j~~~.\eqno(18b)$$  
In terms of these variables, the product  
of $\delta$ functions of Eq.~(10b) can be represented (up to an overall  
constant factor) as
$$\prod_{j=1}^{N^2-1}\delta(f_j)\propto        
 \int dh dH \exp(i {\rm Tr} hA+ i\sum_{j=2}^N H_j {\rm Im} B_{1j})~~~.
 \eqno(19a)$$
The third set are $N-1$ complex numbers $k_j~,~j=2,...,N$, integrated 
along a contour on the real axis with integration measure 
$$dk=\prod_{j=2}^N dk_j/(k_j-i\epsilon)~~~,\eqno(19b)$$           
with infinitesimal positive $\epsilon$.  
These can be used to insert a product of 
step functions $\prod_{j=2}^N \theta({\rm Re} B_{1j})$ into Eq.~(12b),
$$ \prod_{j=2}^N \theta({\rm Re} B_{1j}) \propto \int dk \exp(i\sum_{j=2}^N
k_j {\rm Re} B_{1j})~~~,\eqno(19c)$$
allowing the integrals over the ${\rm Re}B_{1j}$ in Eq.~(12b) to be taken 
from $-\infty$ to $\infty$. 

Defining a matrix $Y$ by 
$Y_{11}=0$; $Y_{ij}=0~,~2\leq i,j \leq N$; 
$Y_{1j}=-Y_{j1}=-{1\over 2}H_j$, the second term in the exponent in  
Eq.~(19a) can be rewritten as $i\sum_{j=2}^N H_j {\rm Im} B_{1j} =
{\rm Tr} YB$, and so an alternative form of Eq.~(19a) is 
$$\prod_{j=1}^{N^2-1}\delta(f_j)\propto        
 \int dh dH \exp(i {\rm Tr} hA+ {\rm Tr}YB)~~~.
 \eqno(19d)$$
Similarly, defining a matrix $Z$ by 
$Z_{11}=0$; $Z_{ij}=0~,~2\leq i,j \leq N$; 
$Z_{1j}=Z_{j1}={1\over 2}ik_j$, the exponent in Eq.~(19c) can be rewritten   
as $\sum_{j=2}^Nik_j{\rm Re}B_{1j}={\rm Tr} ZB$, and so an 
alternative form of Eq.~(19c) is 
$$ \prod_{j=2}^N \theta({\rm Re} B_{1j}) \propto \int dk \exp({\rm Tr} ZB )
~~~.\eqno(19e)$$
          These equations 
allow us to write Eq.~(12b) in terms of the unrestricted measure $dM$, 
and the ghost representation of $\Delta$, as  
$$\eqalign{
{\cal J}=&C \int dM dh dH dk d\omega d\tilde \omega \cr
\times & 
\exp\big(i {\rm Tr} hA+ {\rm Tr} \tilde \omega[\omega,A] 
+ \sum_{j=2}^N(iH_j {\rm Im} B_{1j} +ik_j {\rm Re} B_{1j}
-\tilde \omega_{jj} {\rm Im} [B,\omega]_{1j} ) \big) {\cal G}[\{M\}]~~~\cr
=&C \int dM dh dH dk d\omega d\tilde \omega
\exp\big(i {\rm Tr} hA+ {\rm Tr} \tilde \omega[\omega,A]   
+{\rm Tr} (X+Y+Z)B \big){\cal G}[\{M\}] ~~~,\cr
}\eqno(20)$$
with $C$ an overall constant factor.   The first representation of 
${\cal J}$ in Eq.~(20) will be used to establish a BRST invariance, 
while the second will be used to discuss Ward identities obeyed by the 
matrix-valued correlations.  

We now show that the first integral in Eq.~(20) is   
invariant under the nilpotent BRST transformation 
$$\eqalign{
\delta A=& [A,\omega] \theta~~~,\cr
\delta B=&[B,\omega]\theta~~~,\cr 
\delta M_d=& [M_d,\omega] \theta~,~d=3,...,D~~~,\cr 
\delta \omega=&\omega^2 \theta~~~,\cr 
\delta \tilde \omega_{ij}=&-ih_{ij} \theta~,~i\not=j~~~,\cr
\delta \tilde \omega_{jj}=&-iH_j \theta~,j=2,...,N~~~,\cr
\delta h=&0 ~~~,\cr
\delta H_j=&0~~~,\cr
\delta k_j=&0~~~,\cr
}\eqno(21)$$
with $\theta$ a $c$-number Grassmann parameter. (The part of this  
transformation involving $\omega$ is  
patterned after the BRST 
transformation for the local operator gauge invariant case studied by 
Adler [9].) 
We first remark that since Eq.~(21) has the form of an infinitesimal unitary  
transformation with generator $\omega \theta$ acting on the matrix 
variables $M_d$, the global unitary invariant function ${\cal G}[\{M\}]$ 
and the matrix integration measure $dM$ are both invariant.  
We consider next the terms in the exponent in Eq.~(20).  From Eq.~(21) 
we have 
$$\eqalign{
\delta [A,\omega]=& [\delta A, \omega] + [A,\delta \omega] 
=[[A,\omega]\theta,\omega] +[A,\omega^2 \theta] \cr
=&-(\omega[A,\omega] +[A,\omega] \omega)\theta + [A,\omega^2] \theta 
= -[A,\omega^2] \theta+  [A,\omega^2] \theta =0~~~.\cr
}\eqno(22a)$$
Hence for the terms in the exponent of Eq.~(20) involving $A$, we get 
(using the fact that $A$ is diagonal)  
$$\eqalign{
&\delta( i {\rm Tr} hA+ {\rm Tr} \tilde \omega[\omega,A] ) =  
i {\rm Tr} h\delta A+ {\rm Tr} (\delta\tilde \omega)[\omega,A] \cr  
&=i{\rm Tr} h [A,\omega]\theta + {\rm Tr} (-ih\theta) [\omega,A]=0~~~.\cr
}\eqno(22b)$$
{}For the terms in the exponent of Eq.~(20) involving $B$ but not involving 
the parameters $k_j$, inside 
the summation over $j$ we have  
$$\eqalign{
&\delta(iH_j {\rm Im} B_{1j} 
-\tilde \omega_{jj} {\rm Im} [B,\omega]_{1j} )=
iH_j {\rm Im}\delta B_{1j} 
-(\delta\tilde \omega_{jj}) {\rm Im} [B,\omega]_{1j} \cr
&= iH_j{\rm Im} [B,\omega]_{1j}\theta
+iH_j\theta {\rm Im} [B,\omega]_{1j}= 0~~~,\cr
}\eqno(22c)$$
since $\delta [B,\omega]=0$ by the same argument as in Eq.~(22a).  
So the entire exponent of the first representation in Eq.~(20) is BRST 
invariant, apart from the $k_j{\rm Re} B_{1j}$ terms.  But the shifts in the  
${\rm Re}B_{1j}$ are linear in $\omega$ while not involving $\tilde \omega$.  
Thus (since we shall see shortly that the integration measures 
are invariant),  the shifts in the terms in the exponent involving the 
products $k_j{\rm Re}B_{1j}$ make a vanishing contribution to the 
Grassmann integrals, by an argument similar to that used to justify the 
neglect of $S$ in Eq.~(16a).  

An alternative method of including the step functions, that leads to 
a manifestly BRST invariant integrand, is to include in the exponent 
in the first 
representation of Eq.~(20) an additional term  $-\sum_{j=2}^N \kappa_j
{\rm Re}[B,\omega]_{1j}$, with $\kappa_j$ auxiliary Grassmann parameters 
that are {\it not} integrated over.  This term is linear in $\omega$ but 
does not involve $\tilde \omega$, and so again makes a 
vanishing contribution to the Grassmann integrals in Eq.~(20).  The 
BRST transformation of Eq.~(21) is then augmented by the rule
$\delta \kappa_j =-ik_j \theta$, with the result that the combination 
$ik_j {\rm Re}B_{1j}- \kappa_j {\rm Re} [B,\omega]_{1j}$ is manifestly 
BRST invariant.  

Continuing the BRST analysis, 
since ${\rm Tr}\sigma \tau=-{\rm Tr}\tau \sigma$ for any 
two Grassmann odd grade matrices $\tau$ and $\sigma$, we have 
${\rm Tr}\omega^2=-{\rm Tr}\omega^2=0$, and so the condition that 
$\omega$ should be traceless is preserved by Eq.~(21).  (On the other 
hand, $\omega^2_{11}$ is nonzero even when $\omega_{11}$ is zero, which 
is why we must use a traceless condition, rather than a condition 
$\omega_{11}=0$, for $\omega$.)  Also, letting ${}^*$ denote complex 
conjugation, since 
$$(\omega^2)^*_{ji}= \sum_{\ell} \omega_{j\ell}^*\omega_{\ell i}^* 
=\sum_{\ell} \omega_{\ell j} \omega_{i \ell} 
=-\sum_{\ell} \omega_{i \ell}\omega_{\ell j} =-(\omega^2)_{ij}~~~,
\eqno(22d)$$
the property that $\omega$ is anti-self-adjoint is preserved by Eq.~(21).  
The integration measures $dh$  
and $dH$ are trivially invariant, while the measure $d\tilde \omega$ is 
invariant because $\delta\tilde \omega$ has no dependence on $\tilde \omega$.  
Since 
$$\delta(d\omega_{ij})=d(\delta \omega)_{ij}=d(\omega^2\theta)_{ij}=
\big(\omega d\omega
+(d\omega) \omega\big)_{ij} \theta~~~,\eqno(23a)$$
we have 
$$\delta(d\omega_{ij})=(\omega_{ii}d\omega_{ij}
+d\omega_{ij}\omega_{jj})\theta+....
=d\omega_{ij}(\omega_{jj}-\omega_{ii})\theta+...,~~~\eqno(23b)$$
with $...$ denoting terms that contain only  matrix elements 
$d\omega_{i^{\prime}\,                                                    
j^{\prime}}$ with $(i^{\prime},j^{\prime}) \not= (i,j)$. Hence there is   
no Jacobian contribution from the diagonal terms in the measure $d\omega$, 
while the Jacobian arising from transformation of the off-diagonal terms in 
$d\omega$ differs from unity by a term proportional to 
$$\sum_{i\not=j}(\omega_{jj}-\omega_{ii})\theta=0~~~,\eqno(23c)$$
and so the measure $d\omega$ is also invariant.  Finally, nilpotence 
of the BRST transformation follows from Eq.~(22a), and its analogs with 
$A$ replaced by $B$ or by a general $M_d$, together with 
$$\delta \omega^2=\{\delta \omega,\omega\}=\{\omega^2\theta,\omega\} 
=\omega^2\{\theta,\omega\}=0~~~.\eqno(23d)$$
This completes the demonstration of the BRST transformation for the 
first representation in Eq.~(20).  

The second representation in Eq.~(20) can be used to derive Ward 
identities from unitary-fixed expectations of trace polynomials ${\bf Q}$;   
these Ward identities play a central role in the arguments for an emergent 
quantum theory given in Ref. [3].  
Employing   
the specific unitary fixing of Eq.~(20) in the definition of Eqs.~(8a,b), 
as applied to ${\bf Q} = {\rm Tr}Q$, 
and using the cyclic property of the trace to rewrite  
${\rm Tr}\tilde \omega[\omega,A]$  as ${\rm Tr}\{\tilde \omega,\omega\}A$, we have 
$$\hat Z \langle {\bf Q} \rangle_{\hat{\rm AV}}=
\int dM dh dH dk d\omega d\tilde \omega
\exp\big( {\rm Tr}[ (ih+\{\tilde \omega,\omega\}) A]    
+{\rm Tr} (X+Y+Z)B \big)\exp(-{\bf H}) {\bf Q} ~~~,\eqno(24)$$
with $\hat Z$ given by the expression on the right hand side of Eq.~(24) 
with ${\bf Q}$ replaced by unity.  Ward identities follow from the fact that 
the unrestricted measure $dM$ is invariant under a shift of any matrix 
$M_d$ by a constant $\delta M_d$, which under the  
assumption that surface terms related to the shift vanish, implies  
$$0=\int dM dh dH dk d\omega d\tilde \omega  \delta_{M_d}\Big(
\exp\big( {\rm Tr}[ (ih+\{\tilde \omega,\omega\}) A]  
+{\rm Tr} (X+Y+Z)B \big)\exp(-{\bf H}) {\bf Q}\Big) ~~~.\eqno(25a)$$
When ${\bf H}$ and ${\bf Q}$ are varied with respect to $M_d$, the factor 
$\delta M_d$ can be cyclically permuted to the right in each term of the 
varied trace polynomials, giving the formulas
$$\eqalign{
\delta_{M_d} {\bf H} =& {\rm Tr} {\delta {\bf H} \over \delta M_d} \delta M_d
~~~,\cr
\delta_{M_d} {\bf Q} =& {\rm Tr} {\delta {\bf Q} \over \delta M_d} \delta M_d
~~~,\cr
}\eqno(25b)$$
which [3] {\it define} the variational derivatives of the trace polynomials 
with respect to the operator $M_d$.   Carrying through 
the variations of all terms of Eq.~(25a), and dividing by $\hat Z$, 
we are left with an expression of the form 
$$0={\rm Tr}\langle W_d \rangle_{\hat{\rm AV}} \delta M_d~~~.\eqno(26a)$$
However, since $\delta M_d $ is an arbitrary self-adjoint matrix, the 
vanishing of the real and imaginary parts of Eq.~(26a) implies the 
matrix identity 
$$0= \langle W_d \rangle_{\hat{\rm AV}} ~~~.\eqno(26b)$$
{}For $d=3,...,D$, the variation $\delta_{M_d}$ in Eq.~(25a) acts only 
on the product $\exp(-{\bf H}){\bf Q}$, and we have  
$$W_d={\delta {\bf Q} \over \delta M_d}  
-{\bf Q} {\delta {\bf H} \over \delta M_d}~~~.\eqno(27a)$$ 
However, for $d=1$ and $d=2$, corresponding to $M_1=A$ and  
$M_2=B$, there are additional contributions to the Ward identities 
arising from variations of 
the traces involving $A$ and $B$ in the first exponential on the right hand 
side of Eq.~(25a), which arose from the unitary fixing procedure.  
Explicitly, we have 
$$\eqalign{ 
W_1=& (ih+\{\tilde \omega,\omega\}){\bf Q} 
+{\delta {\bf Q} \over \delta A}-{\bf Q} {\delta {\bf H} \over \delta A}
~~~, \cr 
W_2=&(X+Y+Z) {\bf Q} 
+{\delta {\bf Q} \over \delta B}-{\bf Q} {\delta {\bf H} \over \delta B}
~~~. \cr 
}\eqno(27b)$$
Hence from Eq.~(20) we are able to get explicit forms of all of the 
Ward identities, including those obtained by varying the matrices singled 
out in the unitary invariance fixing.  Note that were we to employ the 
original ensemble average of Eq.~(6a), which has no unitary fixing, in  
deriving the Ward identities, then Eq.~(7b) implies that we 
would only obtain the trace of the matrix relation of Eq.~(26b). In  
other words, unitary fixing is essential for extracting the full content 
of the Ward identities; without it, all nontrivial matrix structure 
is averaged out.

\bigskip
\centerline{\bf Acknowledgments}
This work was supported in part by the Department of Energy under
Grant \#DE--FG02--90ER40542.  We wish to thank Herbert Neuberger and Steven  
Weinberg for useful discussions.  
\vfill\eject
\centerline{\bf References}
\bigskip
\noindent
[1]  E. Br\'ezin and S. R. Wadia, eds., The large N expansion in quantum
field theory and statistical mechanics (World Scientific, Singapore, 1993).
\bigskip 
\noindent
[2] W. Taylor, Rev. Mod. Phys. 73 (2001) 419.
\bigskip
\noindent
[3] S. L. Adler, Statistical dynamics of global unitary invariant matrix 
models as pre-quantum mechanics, hep-th/0206120; S. L. Adler, Quantum 
theory as an emergent phenomenon (Cambridge University Press, 
Cambridge, to be published). 
\bigskip
\noindent
[4] M. L. Mehta, Random matrices (Academic Press, New York, 1967), Chapt. 3.
\bigskip
\noindent
[5] S. Weinberg, The quantum theory of fields, Vol. II (Cambridge University 
Press, Cambridge, 1996), Sec. 15.5.
\bigskip
\noindent
[6]  S. R. Wadia, Integration over a large random matrix by Faddeev-Popov 
method, unpublished Chicago preprint EFI 81/37 (1981), archived at the 
SLAC library; T. Yoneya and H. Itoyama, Nucl. Phys.  B200 (1982) 439; 
S. Coleman, Aspects of symmetry (Cambridge University Press, 
Cambridge, 1989), pp. 394-395.
\bigskip
\noindent
[7] C. Becchi, A. Rouet, and R. Stora, Comm. Math. Phys. 42 (1975) 127; 
in G. Velo and A. S. Wightman, eds., Renormalization theory (Reidel, 
Dordrecht, 1976); Ann. Phys. 98 (1976) 287.  I. V. Tyutin, 
Lebedev Institute preprint N39 (1975).   For an exposition, see S. Weinberg, 
Ref. [5].
\bigskip
\noindent
[8] N. Nakanishi, Prog. Theor. Phys. 35 (1966) 1111; B. Lautrup, Mat. Fys. 
Medd. Kon. Dan. Vid.-Sel. Medd. 35 (1967) 29.  For an exposition, see S. 
Weinberg, Ref. [5].
\bigskip
\noindent
[9] S. L. Adler, J. Math. Phys. 39 (1998) 1723.  When $\omega$ in this  
reference is taken as a constant matrix, the BRST transformation of Eq.~(14) 
of this reference gives the part of Eq.~(21) above relating to 
the $A$ unitary-fixing conditions.  
\bigskip
\noindent
\bigskip
\noindent
\bigskip
\noindent
\bigskip
\noindent
\bigskip
\noindent
\bigskip
\noindent
\bigskip
\noindent
\bigskip
\noindent
\bigskip
\noindent
\bigskip
\noindent
\bigskip
\noindent
\bigskip
\noindent
\bigskip
\noindent
\vfill
\eject
\bigskip
\bye